\documentclass[11pt]{article}
\usepackage{times}
\usepackage{graphics}
\usepackage{color}
\usepackage{amssymb}
\usepackage{amsmath}
\usepackage{ifthen}
\usepackage{calc}
\usepackage{float}

\newtheorem{theorem}{Theorem}[section]

\newtheorem{proposition}[theorem]{Proposition}

\newlength{\Ainlength}
\newlength{\Ainindent}
\newlength{\Aintemp}
\newcommand{\Ain}[1]{\setlength{\Aintemp}{\Ainindent}\addtolength{\Aintemp}{#1\Ainlength} \hspace*{\Aintemp}}

\renewcommand{\vec}[1]{\ensuremath{\mathbf{#1}}}
\newcommand{\data}{\ensuremath{(\vec{y},\vec{w})}}
\newcommand{\err}[2]{\ensuremath{\mathsf{err}^{#1}\ifthenelse{\equal{#2}{}}{}{(#2)}}}
\newcommand{\mean}[2]{\ensuremath{\mathsf{mean}_{#1}\ifthenelse{\equal{#2}{}}{}{(#2)}}}
\newcommand{\Err}[1]{\ensuremath{\mathsf{e}\ifthenelse{\equal{#1}{}}{}{(#1)}}}
\newcommand{\Errj}[1]{\ensuremath{\mathsf{e^\prime}\ifthenelse{\equal{#1}{}}{}{(#1)}}}
\newcommand{\jopt}[1]{\ensuremath{\mathsf{j_{min}}\ifthenelse{\equal{#1}{}}{}{(#1)}}}
\newcommand{\first}[1]{\ensuremath{\mathsf{first\_step}\ifthenelse{\equal{#1}{}}{}{(#1)}}}
\newcommand{\inter}[2]{\ensuremath{[#1\!:\!#2]}}

\floatstyle{plain}
\newfloat{algorithm}{tbp}{algorithm}

\floatname{algorithm}{Algorithm}

\begin{document}

\begin{center}
\textbf{\Large Optimal Reduced Isotonic Regression}
\bigskip
\bigskip

{\large Janis Hardwick ~~and~~ Quentin F. Stout}
\medskip

jphard@umich.edu ~~~~~~~~~ qstout@umich.edu
\smallskip

University of Michigan\\
Ann Arbor, MI
\end{center}

\medskip

\subsubsection*{Abstract}
Isotonic regression is a shape-constrained nonparametric regression in
which the regression is an increasing step function. 
For $n$ data points, the number of steps in the isotonic regression
may be as large as $n$.
As a result, standard isotonic regression has been 
criticized as overfitting the data or making the representation too complicated.
So-called ``reduced'' isotonic regression constrains the outcome to be a specified number of steps $b$, $b \leq n$. 
However, because the previous algorithms for finding the reduced $L_2$ regression took $\Theta(n+bm^2)$ time, where $m$ is the number of steps of the unconstrained isotonic regression, researchers felt that the algorithms were too slow and instead used approximations.
Other researchers had results that were approximations because they used a greedy top-down approach.
Here we give an algorithm to find an exact solution in $\Theta(n+bm)$ time,
and a simpler algorithm taking $\Theta(n+b m \log m)$ time.
These algorithms also determine optimal $k$-means clustering of weighted 1-dimensional data.
\medskip

\noindent
\textbf{Keywords}: reduced isotonic regression, step function, v-optimal histogram,  
piecewise constant approximation, k-means clustering, nonparametric regression

\section{Introduction}

Isotonic regression is an important form of nonparametric regression 
that allows researchers to relax parametric assumptions 
and replace them with a weaker shape constraint.
A real-valued function $f$ is \textit{isotonic} iff for all $x_1, x_2$ in its domain, 
if $x_1 < x_2$ then $f(x_1) \leq f(x_2)$.
In some settings isotonic functions are called monotonic, while in others monotonic is used to indicate either nondecreasing or nonincreasing.
Myriad uses of isotonic regression can be found in citations to the fundamental books of
Barlow et al.~\cite{BarlowetalBook}
and Robertson et al.~\cite{RobertsonWrightDykstra}.
Nonparametric approaches are increasingly important as researchers encounter situations where parametric assumptions are dubious, and as algorithmic improvements make the calculations practical.

\begin{figure}

\begin{minipage}[b]{2.05in}
\begin{center}
\resizebox{1.85in}{!}{\includegraphics{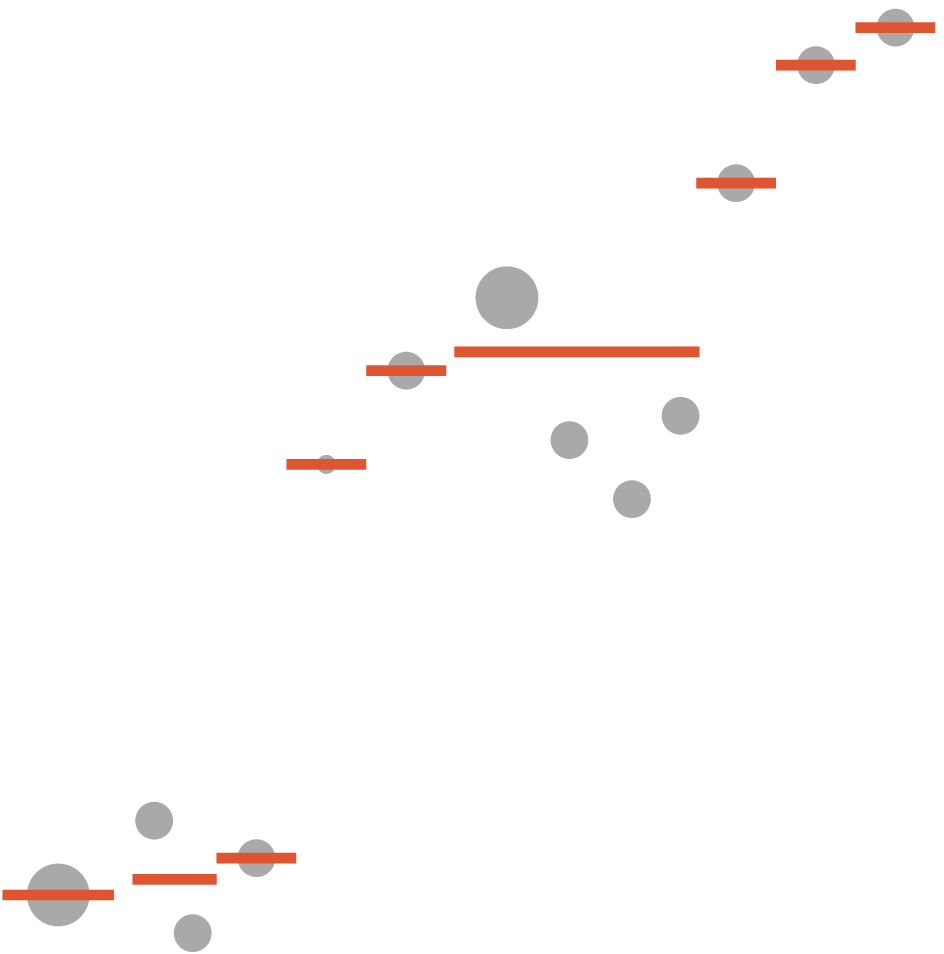}}
\smallskip

a) isotonic regression
\end{center}
\end{minipage}
\hspace{0.15in}
\begin{minipage}[b]{2.05in}
\begin{center}
\resizebox{1.85in}{!}{\includegraphics{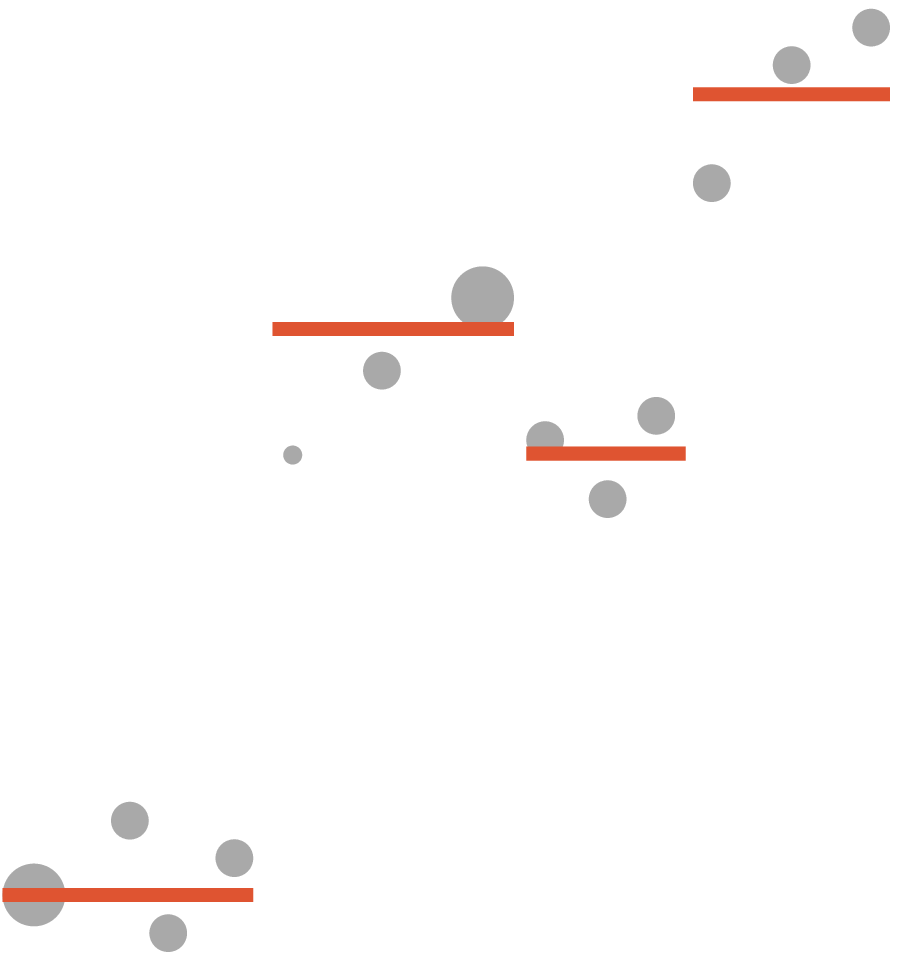}}
\smallskip

b) 4-step regression
\end{center}
\end{minipage}
\hspace{0.15in} 
\begin{minipage}[b]{2.05in}
\begin{center}
\resizebox{1.85in}{!}{\includegraphics{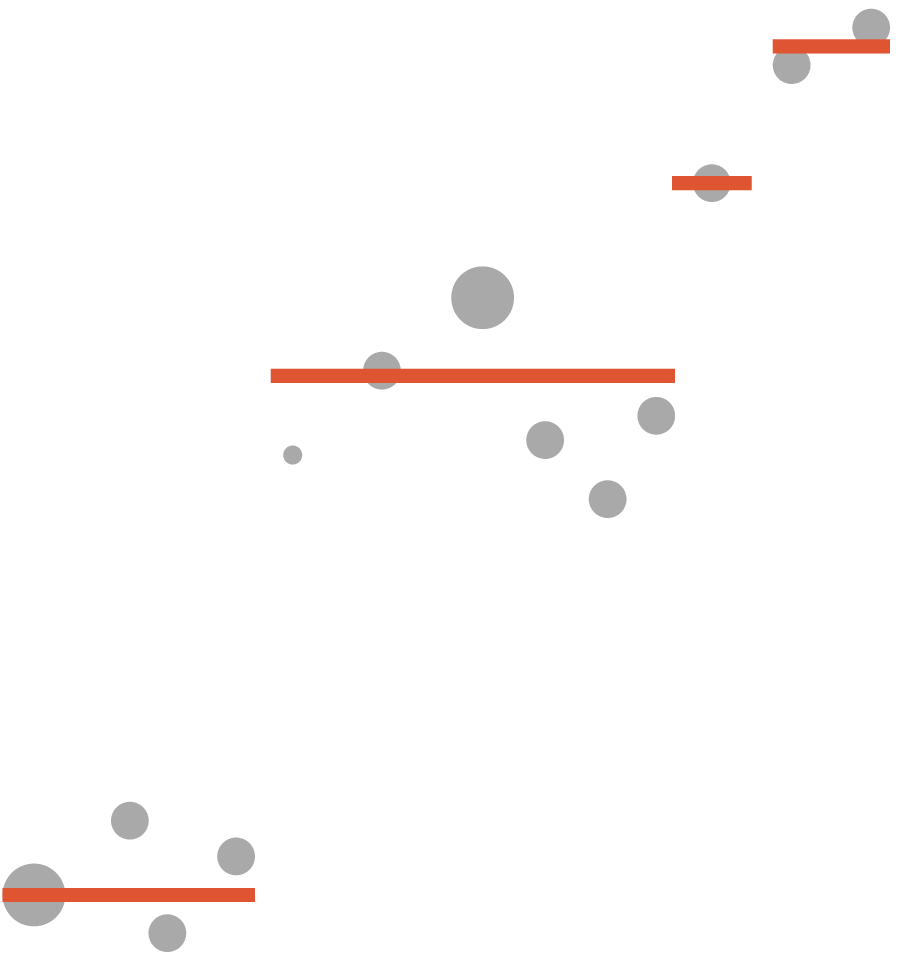}}
\smallskip

c) 4-step reduced isotonic
\end{center}
\end{minipage}

\caption{Stepwise regressions, size indicates weight}\label{fig:Regressions}
\hrulefill
\end{figure}

Isotonic regression is useful for situations in which the independent variable
has an ordering but no natural metric, such as S $<$ M $<$ L $<$ XL clothing sizes.
Since the only important property of the domain is its ordering,
we assume that it is the integers $1 \ldots n$ for some $n$, and use
\inter{i}{j}, $1 \leq i \leq j \leq n$ to denote the range $i \ldots j$.
By \textit{weighted values} \data\ on \inter{1}{n}, we mean values $(y_i,w_i)$,
$i \in \inter{1}{n}$, where the $y$ values are arbitrary real numbers
and the $w$ values (the weights)
are nonnegative real numbers.
Given weighted values \data\ and a real-valued function $f$ on \inter{1}{n}, 
the $L_p$ regression or approximation error of $f$ is
$$
\begin{array}{ll}
  \left(\sum_{i=1}^n w_i |y_i -f(i)|^p\right)^{1/p} & 1 \leq p < \infty \medskip \\
  \max_{i=1}^n w_i |y_i - f(i)|                     & p = \infty
\end{array}
$$
An $L_p$ \textit{isotonic regression} is an isotonic function that minimizes
the $L_p$ error among all isotonic functions. 
Figure~\ref{fig:Regressions}~a) gives an example of an isotonic regression. 
Because researchers from varying fields often use different expressions for a single concept, 
we use the terms {\it regression} and {\it approximation} interchangeably. We identify approximations that are not optimal regressions as {\it sub-optimal} approximations.

Isotonic regressions are  step functions for which   
the number of steps is determined by the data.
In certain cases there is criticism that such functions can overfit the data~\cite{NiculescuMizilCaruana,SalantiUlmReduced,SchellSingh}
or produce a result with too many steps~\cite{HaiminenetalReducedUnimodal}.
Consequently, some researchers utilize isotonic regressions that
restrict the number of steps.  
Schell and Singh~ \cite{SchellSingh} have referred to such functions as
\textit{reduced} isotonic regressions.

Restricting the number of steps is a central issue in approximation by step functions. It arises in settings such as databases and variable width
histogramming~\cite{HalimetalFastHistogram,IoannidisIncreasingHist,PoosalaetalHistogramSurvey}, segmentation of time series and genomic data~\cite{HimbergetalTimeSeries,JacobetalFuzzySegmentation,TerziSegmentation},
homogenization~\cite{FisherHomogeneity}
and piecewise constant approximations~\cite{MaysterLopezStep}.

A function $f$ is an \textit{optimal $L_p$ $b$-step approximation}, $1 \leq b \leq n$,
iff it minimizes the $L_p$ error over all functions with $b$ steps. 
Here we are primarily concerned with computing $L_2$ $b$-step reduced isotonic regressions, where
a function 
 $f$ is an \textit{optimal $L_p$ $b$-step reduced isotonic regression}, 
$b=1,\ldots,m \leq n$,
iff it minimizes the $L_p$ error over all isotonic functions having $b$ steps. 
Figure~\ref{fig:Regressions} gives examples of $b$-step regression and $b$-step reduced isotonic regression.
Optimal $b$-step approximations and $b$-step reduced isotonic regressions are not always unique.
For example, with unweighted values 1, 2, 3 on \inter{1}{3} and $b=2$, 
for any $p$ the function
which is 1.5 on \inter{1}{2} and 3 at 3 is optimal, as is the function which is 1 at
1 and 2.5 on \inter{2}{3}.

In 1958 Fisher~\cite{FisherHomogeneity} gave a simple algorithm
for determining an optimal $b$-step $L_2$ regression in $\Theta(b n^2)$ time
(this is shown in Algorithm~\ref{alg:FisherDP}).
His algorithm can be easily modified to determine an optimal $b$-step $L_2$ reduced isotonic regression in the same time bounds.
His algorithm has been widely used and rediscovered, and often falsely attributed to Bellman.
However, for many researchers the quadratic time in $n$ makes it too slow for their
applications~\cite{HaiminenetalReducedUnimodal,HalimetalFastHistogram,HimbergetalTimeSeries,JacobetalFuzzySegmentation,TerziSegmentation}.
Thus most previous work utilizing reduced isotonic regression used sub-optimal approximations, with the exception of an algorithm due
to Haiminen, Gionis and Laasonen~\cite{HaiminenetalReducedUnimodal}. 
Their algorithm for the $L_2$ metric takes $\Theta(n + bm^2)$ time, where $m$ is the number of pieces of the unrestricted isotonic regression.
(To lessen confusion, we use ``pieces'' to refer to the steps of the unrestricted isotonic regression.)
However, even with this reduction in time they then developed an approximation algorithm based on a greedy heuristic.

In Section~\ref{sec:Reduced} we decrease the time to find the optimal $b$-step $L_2$ reduced isotonic regression to $\Theta(n + b m)$, using an
algorithm in Section~\ref{sec:Isotonic} for the special case in which the values are themselves isotonic.
A simpler algorithm, taking $\Theta(n + b m \log m)$ time, is also given.
These algorithms should be fast enough to eliminate the need for approximations, even for very large data sets.

Since we are only looking for optimal approximations, we often omit ``optimal''.

\section{Approximation by Step Functions}\label{sec:ArbitraryStep}

A real-valued function $f$ on \inter{1}{n} is a 
\textit{$b$-step function}, $1 \leq b \leq n$,  
iff there are indices 
$j_0 = 0 < j_1  \ldots < j_b=n$ 
and real values
$C_k$, $k \in \inter{1}{b}$,
such that $f(x_i) = C_k$ for $i \in \inter{j_{k\!-\!1}+\!1}{j_k}$. 
If $f$ is isotonic then $C_1 \leq C_2 \ldots \leq C_b$.
An approximation with fewer than $b$ steps can be converted to a $b$-step
approximation by merely subdividing steps, and thus we do not differentiate
between ``$b$ steps'' and ``no more than $b$ steps''.

Let \mean{p}{i,j} denote an $L_p$ mean of the weighted values on \inter{i}{j}.
For $1 \leq p < \infty$, an optimal $L_p$ step function
has the property that $C_k = \mean{p}{j_{k\!-\!1}+\!1,\,j_{k}}$.
Since we are only concerned with optimal approximations, whenever
a function has a step \inter{i}{j}, then its value on that step is \mean{p}{i,j}.
Let \err{p}{i,j} denote the $p^\mathrm{th}$ power of the $L_p$ error of the step \inter{i}{j}.
Minimizing the sum of the \err{p}{} values is the same as minimizing the
$L_p$ approximation error and thus from now on only the \err{p}{} values will be used.


\subsection{Arbitrary Data}\label{sec:ArbitraryData}

Fisher's~\cite{FisherHomogeneity} dynamic programming approach to determining
an optimal $L_p$ $b$-step approximation for $1 \leq p < \infty$
is based on the observation that
if $f$ is an optimal $b$-step approximation of the data, with a first step of \inter{1}{j},
then $f$ is an optimal $(b\!-\!1)$-step approximation of
the data on \inter{j\!+\!1}{n}.
This is obvious since if it were not optimal then replacing it with
an optimal $(b\!-\!1)$-step approximation would reduce the error.
Let \Err{i,c} denote the sum of the \err{p}{} values of the steps of an
optimal $c$-step approximation on \inter{i}{n}, and let 
\Errj{i,j,c} denote the sums of the \err{p}{} values of the steps of a
$c$-step approximation on \inter{i}{n} which is optimal among $c$-step approximations
where the first step is \inter{i}{j}.
Fisher's observation yields the equations:
\begin{eqnarray}
 \Errj{i,j,c} & = & \err{p}{i,j} + \Err{j\!+\!1,c\!-\!1}   \label{eqn:Errj}\\
 \Err{i,c} & = & \min\{\Errj{i,j,c}: i \leq j \leq n-c+1\}
\end{eqnarray}
By storing the $j$ that minimizes \Err{i,c} in \jopt{i,c},
in $\Theta(n)$ time
one can generate the optimal approximation after the dynamic programming has completed.
This leads to Algorithm~\ref{alg:FisherDP}.
The time is $\Theta(bn^2)$ plus the time to compute the 
$\Theta(n^2)$ \err{p}{} values.
For $L_\infty$, $\Errj{i,j,c} = \max\{\err{\infty}{i,j},~ \Err{j\!+\!1,c\!-\!1}\}$.

Fisher's algorithm can be modified to determine the $b$-step reduced isotonic regression in the same time bounds.
The lines
\medskip

\noindent
\setlength{\Ainindent}{0.5in}
\Ain{1}    $\mathsf{for~ i=1 ~to~ n-c+1}$\\
\Ain{2}       $\mathsf{\Err{i,c} = min\{\Errj{i,j,c}: i \leq j \leq n-c+1\}}$
\medskip

\noindent
should be replaced by
\medskip

\noindent
\setlength{\Ainindent}{0.0in}
\Ain{0} \hspace*{-0.05in}$\mathsf{for~ i=1 ~to~ n-1}$\\
\Ain{0} \hspace*{0.11in} $\mathsf{\Err{i,c} = min \bigl\{\err{p}{i,n},~\min\{\Errj{i,j,c}: i \leq j \leq n\!-\!1,~  mean_p(i,j) \leq mean_p(j\!+\!1,\,\jopt{j\!+\!1,c\!-\!1})\,\}\, \bigr\}}$
\medskip

\noindent
Including the $\err{p}{i,n}$ term, and changing the upper bound on $\mathsf{i}$, is necessary so that, say, for unweighted data 3, 2, 1, the $L_2$ 2-step reduced isotonic regression is correctly determined to be 2, 2, 2.
Using either 3, or 3, 2, as the initial step would involve a second step that was lower, and hence the solution has only 1 step.

Throughout, the values of \Err{} and \jopt{} are stored in 2-dimensional arrays, while \Errj{} is evaluated as a function, not stored as a 3-dimensional array.
To evaluate \err{2}{}, once the scan values $\sum_{j=1}^i w_j y_j$,\, 
$\sum_{j=1}^i w_j y_j^2$,\, and $\sum_{j=1}^i w_i$ have been determined for all $i \in \inter{1}{n}$,
each \err{2}{} value can then be computed in unit time.

\begin{algorithm}  
\setlength{\Ainindent}{1.0in}

\Ain{0} $\mathsf{for~ i = 1~ to~ n}$\\
\Ain{1}    $\mathsf{e(i,1) = \err{p}{i,n};~~~ \jopt{\mathsf{i,1}} = i}$\\
\Ain{0} $\mathsf{for~ c= 2~ to~ b}$\\
\Ain{1}    $\mathsf{for~ i=1 ~to~ n-c+1}$\\
\Ain{2}       $\mathsf{\Err{i,c} = min\{\Errj{i,j,c}: i \leq j \leq n-c+1\}~~~~\{\Errj{}~ is~ defined~ in~ (\ref{eqn:Errj})\}}$\\
\Ain{3}           $\mathsf{\{record~ minimizing~ j~ in~ \jopt{\mathsf{i,c}}\}}$\\
\Ain{1} $\mathsf{end~ for~ i}$\\
\Ain{0} $\mathsf{end~ for~ c}$\\
\Ain{0} $\mathsf{generate~ the~ approximation~ using~ \jopt{}~ and ~\mean{p}{}}$

\vspace*{0.3in} 
\hrulefill
\vspace*{-0.4in}

\caption{Fisher's algorithm for optimal $L_p$ $b$-step approximation of arbitrary data, $1 \leq p \leq \infty$} \label{alg:FisherDP}

\end{algorithm}

\subsection{Isotonic Data}\label{sec:Isotonic}

Reducing the time of Algorithm~\ref{alg:FisherDP}
requires reducing the number of \err{p}{} values referenced.
It is not known how to do this for arbitrary data, but isotonic data has some special properties.
We give two algorithms: Algorithm~\ref{alg:monotonic} is simpler than Algorithm~\ref{alg:totallymonotonic},
but, in O-notation, slower by a logarithmic factor.
It is likely that many will prefer Algorithm~\ref{alg:monotonic} over Algorithm~\ref{alg:totallymonotonic}.
Algorithm~\ref{alg:monotonic} is given in Section~\ref{sec:monotonic}, and Algorithm~\ref{alg:totallymonotonic} is in Section~\ref{sec:totallymonotonic}.

For isotonic data, the fact that values are nondecreasing allows one to make
inferences concerning the means of intervals.
For example, the $L_p$ mean of the weighted values on \inter{i}{j} is no larger than 
that of the values on \inter{i\!+\!1}{j}.
Further, for any $1 < i \leq j < n$,
$\err{p}{i,j+1} - \err{p}{i,j} \geq \err{p}{i+1,j+1} - \err{p}{i+1,j}$.
That is, if we consider the increase in error of adding $(x_{j+1},w_{j+1})$ to the step \inter{i}{j},
this is greater than the increase when adding it to the step \inter{i+1}{j}.
This is true because the monotonicity insures that $x_{j+1}$ is at least as large as the mean on \inter{i+1}{j}, which has a mean not more than that of \inter{i}{j}, and the total weight of \inter{i}{j} is greater than the total weight of \inter{i+1}{j}.
When the values are not isotonic then this inequality may not hold.

Letting $M(i,j) = \err{p}{i,j}$, this can be rewritten as
\begin{equation}
M(i,j\!+\!1) + M(i\!+\!1,j) \geq M(i,j) + M(i\!+\!1,j\!+\!1) \label{eqn:Monge1}
\end{equation}
for all $1 \leq i < j < n$ and $1 \leq p \leq \infty$.
This is known as the \textit{Monge property}, and $M$ is known as a Monge matrix
(typically the Monge property has the inequality in the opposite order and is applied to maximization, not minimizing).

If \jopt{i} denotes the smallest $j$ such that $M(i,j)$ is a minimal value in row $i$ of $M$,
then the Monge property implies that for any $i < i^\prime$, ~$\jopt{i} \leq \jopt{i^\prime}$,
i.e., \jopt{} is isotonic.
This property is typically called \textit{monotonicity}.
If we define $M(i,j)=\infty$ when $j<i$ then $M$ satisfies~(\ref{eqn:Monge1}) for all $i$ and $j$.
Iteratively combining this inequality over adjacent elements shows that it holds much more widely,
in that for all $1 \leq i_1 < i_2 \leq n$ and $1 \leq j_1 < j_2 \leq n$,
\begin{equation}
M(i_1,j_2) + M(i_2,j_1) \geq M(i_1,j_1) + M(i_2,j_2)  \label{eqn:Monge2}
\end{equation}
Thus all submatricies of a Monge matrix are Monge, where a submatrix can be formed from an arbitrary set of rows and an arbitrary set of columns and the number of rows need not equal the number of columns.
Since all submatricies are Monge, all are monotonic. 
This property is called \textit{total monotonicity}.
There are monotonic matrices that are not totally monotonic and
totally monotonic matrices that aren't Monge.

The fact that $M$ is a Monge matrix implies that $M^c$ is a Monge matrix, for $c > 1$,
where $M^c(i,j) = \Errj{i,j,c}$.
This is because 
\begin{eqnarray*}
M^c(i,j\!+\!1)+M^c(\!i+\!1,j) & = & M(i,j\!+\!1) + \Err{j\!+\!1,c\!-\!1} + M(i\!+\!1,j) + \Err{j\!+\!2,c\!-\!1} \\
M^c(i\!+\!1,j\!+\!1)+M^c(i,j) & = & M(i\!+\!1,j\!+\!1) + \Err{j\!+\!2,c\!-\!1} + M(i,j) + \Err{j\!+\!1,c\!-\!1}
\end{eqnarray*}
Algorithm~\ref{alg:monotonic}, in Section~\ref{sec:monotonic}, exploits the monotonicity of $M^c$ and Algorithm~\ref{alg:totallymonotonic}, in Section~\ref{sec:totallymonotonic}, exploits its total monotonicity.
We will show

\begin{theorem} \label{thm:isotonic}
Given $n$ isotonic weighted values \data\ and number of steps $b \leq n$,
Algorithm~\ref{alg:monotonic} finds an optimal $L_2$ $b$-step
approximation (hence an optimal $L_2$ $b$-step reduced isotonic regression),
in $\Theta(b n \log n)$ time, and Algorithm~\ref{alg:totallymonotonic} finds one in $\Theta(bn)$ time.
$\Box$
\end{theorem}

\subsection{Using Monotonicity} \label{sec:monotonic}

\begin{figure}
\centerline{\resizebox{3.5in}{!}{\includegraphics{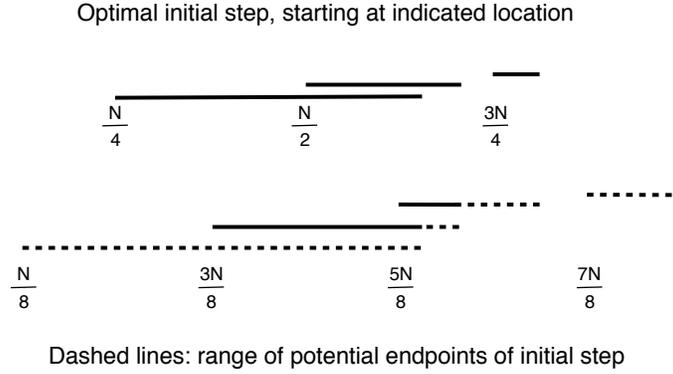}}}
\caption{Possible endpoints of odd multiples of 1/8}  \label{fig:linearscan}
\hrulefill
\end{figure}

Let \jopt{i,b} denote the smallest $j$ such that $\Errj{i,j,b} = \Err{i,b}$.
As noted, \jopt{\cdot,b} is an isotonic function.
This fact can be used to efficiently compute \Err{\cdot,b} and \jopt{\cdot,b}
from the values of \Err{\cdot,b-1} and \jopt{\cdot,b-1}.
Figure~\ref{fig:linearscan} shows an intermediate stage of the calculations for 
a single stage.
The optimal first step for each multiple of 1/4 has been computed and now
the first step for each odd multiple of 1/8 needs to be determined.
For each of these, the possible values of the endpoint of the optimal first step 
are the range indicated by the 
dashed lines with the solid line indicating the part that any optimal first
step must include.

This observation forms the basis of Algorithm~\ref{alg:monotonic}.
Compared to Fisher's algorithm, for fixed $c$, the order in which \Err{i,c} values are determined is changed, as is the range of $j$ values used to compute each value.

\begin{algorithm} 
\setlength{\Ainindent}{1.5in}
\Ain{0} $\mathsf{j\_start \ldots j\_end:~ range~ of~ possible~ endpoints}$\\

\vspace*{-0.08in}
\Ain{0} $\mathsf{for~ i = 1~to~n~ do}$\\
\Ain{1} $\mathsf{   \Err{i,1} = \err{p}{i,n};~~~ \jopt{i,1} = n}$\\
\Ain{0} $\mathsf{for~ c= 2~ to~ b~ do}$\\
\Ain{1} $\mathsf{   for~ level = \lfloor \log_2 (n\!-\!c\!+\!1) \rfloor ~downto~ 0 ~do}$\\
\Ain{2} $\mathsf{      for~ i = 2^{level} ~to~ n-c+1 ~by~ 2^{level\!+\!1}~ do}$\\
\Ain{3} $\mathsf{         if~ i = 2^{level} ~then~ j\_start = j}$\\
\Ain{4} $\mathsf{            else~ j\_start = max\{i, \jopt{i-2^k,c}\}}$\\
\Ain{3} $\mathsf{         if~ i+2^{level} > n-c+1 ~then~ j\_end = n-c+1}$\\
\Ain{4} $\mathsf{            else~ j\_end = \jopt{i+2^{level},c}}$\\
\Ain{3} $\mathsf{         \Err{i,c} = min\{ \Errj{i,j,c}: j\_start \leq j \leq j\_end \}}$\\
\Ain{4} $\mathsf{         \{store~ largest~ minimizing~ j~ in~ \jopt{i,c}\}}$\\
\Ain{2} $\mathsf{      end~ for~ i}$\\
\Ain{1} $\mathsf{   end~ for~ level}$\\
\Ain{0} $\mathsf{end~ for~ c}$\\
\Ain{0} $\mathsf{generate~ the~ approximation~ using~ \jopt{}~ and ~\mean{p}{}}$

\vspace*{0.3in} 
\hrulefill
\vspace*{-0.4in}

\caption{$b$-step $L_p$ approximation of isotonic data, using monotonicity}  \label{alg:monotonic}

\end{algorithm}

\begin{proposition} \label{prop:monotonic}
Given $n$ isotonic weighted values \data\ and number of steps $b \leq n$,
Algorithm~\ref{alg:monotonic} finds an optimal $b$-step
$L_2$ approximation in $\Theta(b n \log n)$ time.
\end{proposition}
\noindent 
\textit{Proof:\,}
Suppose that \Err{\cdot,c} and \jopt{\cdot,c} have been determined for 
$i_1 < i_2 \ldots < i_k$.
Let $\ell_o \ldots \ell_k$ be such that $\ell_0 < i_1 < \ell_1 < i_2 \ldots < i_k <\ell_k$.
To determine \Err{\cdot,c} and \jopt{\cdot,c} for the $\ell$ values, note that since \jopt{\cdot,c} is isotonic
then $\jopt{\ell_0,c} \in \inter{\ell_0}{\jopt{i_1,c}}$, 
$\jopt{\ell_1,c} \in \inter{\max\{\ell_1,\jopt{i_1,c}\}}{\jopt{i_2,c}}$, \ldots, and 
$\jopt{\ell_k,c} \in \inter{\max\{\ell_k, \jopt{i_k,c}\}}{n\!-\!c\!+\!1}$.
Thus, to determine \Err{\ell_0,c} and \jopt{\ell_0,c} we only need to evaluate
\Errj{\ell_0,j,c} for $j \in \inter{\ell_0}{\jopt{i_1,c}}$;
to determine \Err{\ell_1,c} and \jopt{\ell_1,c} we only need to evaluate
\Errj{\ell_1,j,c} for $j \in \inter{\max\{\ell_1,\jopt{i_1,c}\}}{\jopt{i_2,b}}$;
and so forth; i.e., we need at most $n + k$ total evaluations.
In Figure~\ref{fig:linearscan}, this corresponds to the fact that the dashed lines
can overlap only at endpoints.
In $1 +\lfloor \log_2 n \rfloor$ iterations
all values of \Err{\cdot,c} and \jopt{\cdot,c} can be determined.
This gives Algorithm~\ref{alg:monotonic}.

To complete the proof we need to show that each iteration
of the ``$\mathsf{for~level}$'' loop
can be completed in $\Theta(n)$ time.
The $\mathsf{j\_start}$ and $\mathsf{j\_end}$ values that control the number of $j$ values examined guarantee that, over all $i$ values in in ``$\mathsf{for~level}$'' loop, a given $j$ value is used at most twice.
$\Box$

\subsection{Using Total Monotonicity} \label{sec:totallymonotonic}

The fact that $M^c$ is totally mononotonic can be used to further reduce the total number of $j$ values examined.
Algorithm~\ref{alg:totallymonotonic} replaces
\smallskip

\centerline{$\mathsf{\Err{i,c} = min\{ \Errj{i,j,c}: j\_start \leq j \leq j\_end \}}$}
\medskip

\noindent in Algorithm~\ref{alg:monotonic} with a while loop over a smaller set of $j$ values, reducing the worst-case total number used at level $k$ from $n-2^k+1$ to $\lfloor n/2^k \rfloor$.
These $j$ values are determined in Algorithm~\ref{alg:SMAWK}.
The approach used is known as the SMAWK algorithm, an anagram of the initials of the authors of~\cite{SMAWK}.
It is likely that most readers are unfamiliar with SMAWK, and some might prefer to just view Algorithm~\ref{alg:SMAWK} as a black box having the properties that for every $c$:
\begin{itemize}
\item for any level $k$ and any $i$ for which \jopt{i} is determined at level $k$, $\mathsf{jvalues}(k,\cdot)$ contains \jopt{i},
\item the total number of $j$ values returned over all levels is $O(n)$,
\item $\mathsf{determine\_jvalues}$ takes $\Theta(n)$ time.
\end{itemize}
The pseudo-code given in Algorithm~\ref{alg:SMAWK} is quite explicit, suitable for efficient implementation in any language.
It converts the recursive list-based description in~\cite{SMAWK} to an iterative array-based one.
Mention of eliminating columns, creating submatrices, etc., is merely symbolic since there aren't any real matrices:
they are just conceptual representations of calculating \Errj{i,j,c} values.
The only arrays being used are to store $j$ values.

To see how the SMAWK algorithm works, let $M$ denote an arbitrary totally monotonic matrix.
The algorithm starts with a list of columns $J$ ($\mathsf{jvalues}$), and a subset of them are moved to $K$ and kept, with the remaining ones deleted.
The final set of values in $K$ will be the ones returned by $\mathsf{determine\_jvalues}$.
When a column $m$ is deleted from $J$ and not put into $K$
it is guaranteed that for all rows $i$, $m \neq \jopt{i}$.
The guarantees come about by exploiting two facts implied by the general Monge property~(\ref{eqn:Monge2}):
for the $2 \times 2$ submatrix with columns $\alpha < \beta$ and rows $\delta < \epsilon$,
\begin{enumerate}
\item[a)] if $\beta$ is the minimal location in row $\delta$, i.e., $M(\delta,\alpha) > M(\delta,\beta)$, then it is the minimal location in row $\epsilon$, and hence in $M$ $\alpha$ is not the minimal location in any row $\geq \delta$
\item[b)] if $\alpha$ is the minimal location in row $\epsilon$, i.e., $M(\epsilon,\alpha) \leq M(\epsilon,\beta)$, then it is the minimal location in row $\delta$, and hence in $M$ $\beta$ is not the minimal location in any row $\leq \epsilon$
\end{enumerate}

\begin{figure}
\begin{center}
\resizebox{5.0in}{!}{\includegraphics{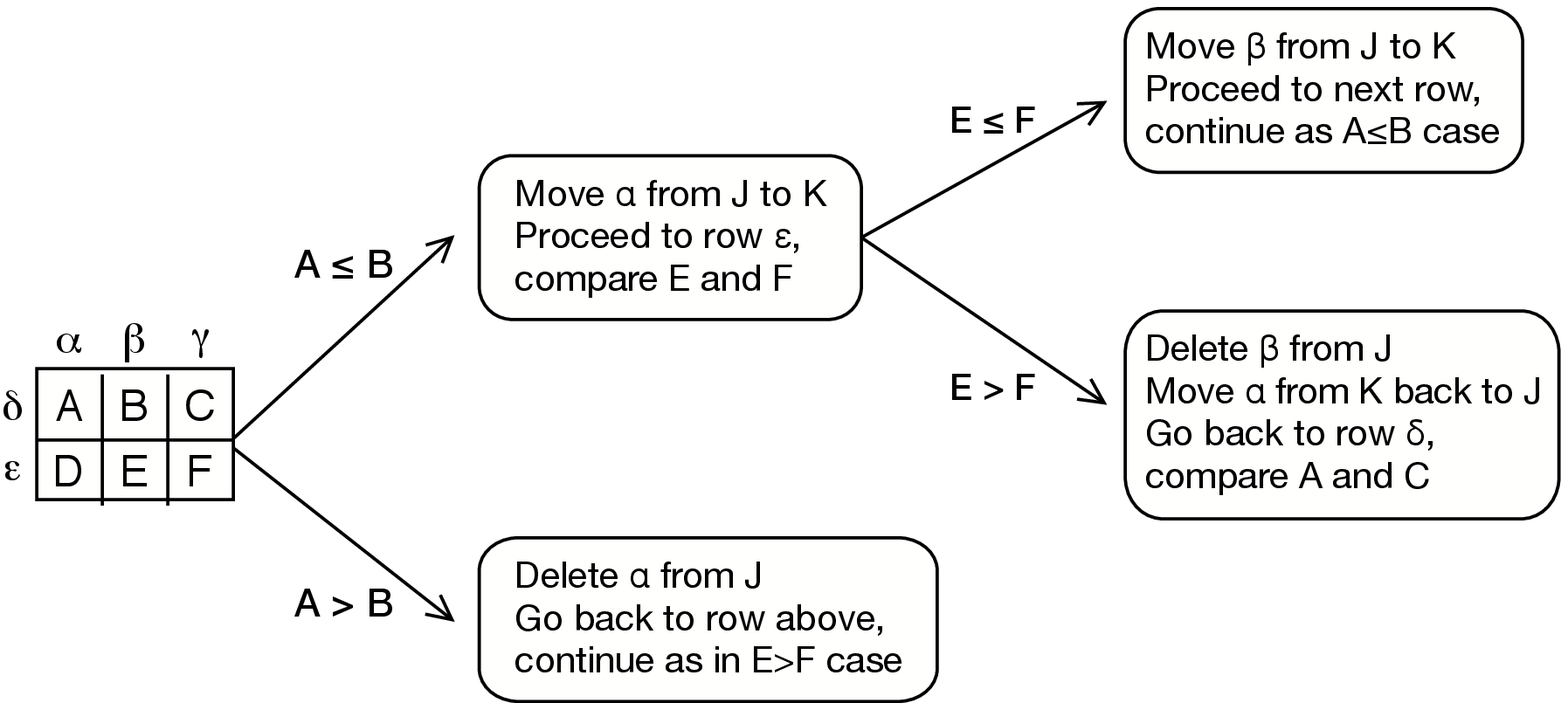}}
\vspace{0.25in}

Currently at row $\delta$, comparing A and B.\\
$\alpha$, $\beta$, $\gamma$ are the first 3 columns remaining in $J$;~~ $\delta$, $\epsilon$ are consecutive rows in the submatrix
\end{center}
\caption{An intermediate step of the SMAWK algorithm}  \label{fig:SMAWKflowchart}
\hrulefill
\end{figure}

At any step in the algorithm two adjacent entries of $M$ are being compared, where they are in the same row and the first two columns ($j$ values) remaining in $J$.
For every row above the current row, one column has been moved into $K$.
Suppose the algorithm is comparing $A$ and $B$ in Figure~\ref{fig:SMAWKflowchart}.
If $A\leq B$ then it might be that $\alpha = \jopt{\delta}$, 
and hence $\alpha$ is moved from $J$ to $K$.
Note that $\alpha$ might also be \jopt{} for some rows above and below $\delta$.
Relative to row $\delta$, column $\beta$ does not need to be kept.
Further, for any row above $\delta$, Monge property~b) shows that $\beta$ is not needed there either.
However, it might be needed for lower rows, so the algorithm proceeds to the next row, $\epsilon$, and compares E and F.
If $E\leq F$ then $\beta$ is moved to $K$ and the algorithm proceeds to the next row.
However, if $E > F$ then $\beta$ is not needed for row $\epsilon$, and Monge property~a) shows that it is not needed for any row below.
Therefore $\beta$ can be deleted from $J$, which in the implementation is done by merely incrementing $\mathsf{next\_j\_index}$.
Deleting $\beta$ condenses the submatrix in Figure~\ref{fig:SMAWKflowchart} to the entries A, C, D, and F.
It might be that $A > C$, so the algorithm moves $\alpha$ from $K$ back to $J$ and goes back to row $\delta$, comparing $A$ and $C$.
If A $\leq$ C then $\alpha$ is put back in $K$ and the algorithm goes to the next row ($\epsilon$),
otherwise it is removed from $J$ and the algorithm backs up another row, etc.
If E\,$=\infty$, i.e., $\beta < \epsilon$, then we treat it as E\,$>$\,F even if F\,$=\infty$.

If $\epsilon$ is the last row, if $E \leq F$ then $\gamma$ can be deleted from $J$ since there are no lower rows for which $\gamma$ might need to be kept.
Combining this with the rule that if $E > F$ then $\beta$ is deleted and the algorithm goes back a row shows that if the last row is reached then all of the remaining columns are examined.
Whether it occurs in the last row or earlier, eventually there is only 1 column left,
which should be kept.
Any row results in one column being moved to $K$, or is a row after the row in which the last column is reached, and hence $|K|$ is no more than the number of rows.
Further, the time required is $\Theta(|J|)$.

To initialize, for level 0, which corresponds to all rows, all columns are kept, i.e., $\mathsf{jvalues(0,k)=k}$ for $1 \leq \mathsf{k} \leq n$.
One could apply the above reduction for level 0, but it isn't required for the time analysis nor correctness, and it slightly simplifies the implementation.
At any level $m$ above 0, the process is applied to the submatrix consisting of every second row of the submatrix used for level $m-1$, i.e., to rows that are multiples of $2^m$.
The initial $J$ for level $m$ is $\mathsf{jvalues(m\!-\!1,\,1:num\_jvalues(m\!-\!1))}$.

\begin{algorithm} 
\setlength{\Ainindent}{0.5in}
\Ain{0} $\mathsf{integer~ array~ jvalues(0\!:\!\lfloor \log_2 \rfloor, 1\!:\!n),~num\_jvalues(0\!:\!\lfloor \log_2 n \rfloor)}$\\

\vspace*{-0.08in}
\Ain{0} $\mathsf{for~ i = 1~to~n~ do}$\\
\Ain{1} $\mathsf{   \Err{i,1} = \err{p}{i,n};~~~ \jopt{i,1} = n}$\\
\Ain{0} $\mathsf{for~ c= 2~ to~ b~ do}$\\
\Ain{1} $\mathsf{   determine\_jvalues(jvalues,num\_jvalues,c)~~~ \{see~ Algorithm~\ref{alg:SMAWK}\}}$\\
\Ain{1} $\mathsf{   for~ level = \lfloor \log_2 (n\!-\!c\!+\!1) \rfloor ~downto~ 0 ~do}$\\
\Ain{2} $\mathsf{      for~ i = 2^{level} ~to~ n-c+1 ~by~ 2^{level\!+\!1}~ do}$\\
\Ain{3} $\mathsf{         if~ i = 2^{level} ~then~ j\_start = i;~~ j\_index = 1}$\\
\Ain{4} $\mathsf{            else~ j\_start = max\{i, \jopt{i-2^{level},c}\}}$\\
\Ain{3} $\mathsf{         if~ i+2^{level} > n-c+1 ~then~ j\_end = n-c+1}$\\
\Ain{4} $\mathsf{            else~ j\_end = \jopt{i+2^{level},c}}$\\
\Ain{3} $\mathsf{         \Err{i,c} = \infty}$\\
\Ain{3} $\mathsf{         while~ (j\_index \leq num\_jvalues(level)) \wedge (jvalues(level,j\_index) \leq j\_end) ~do}$\\
\Ain{4} $\mathsf{            j=jvalues(level,j\_index)}$\\
\Ain{4} $\mathsf{            if~ (j \geq j\_start) \wedge (\Errj{i,j,c} < \Err{i,c}) ~then}$\\
\Ain{5} $\mathsf{               \Err{i,c} = \Errj{i,j,c};~~ \jopt{i,c}=j}$\\
\Ain{4} $\mathsf{            j\_index = j\_index + 1}$\\
\Ain{3} $\mathsf{         end~ while}$\\
\Ain{3} $\mathsf{         j\_index=j\_index-1}$\\
\Ain{2} $\mathsf{      end~ for~ i}$\\
\Ain{1} $\mathsf{   end~ for~ level}$\\
\Ain{0} $\mathsf{end~ for~ c}$\\
\Ain{0} $\mathsf{generate~ the~ approximation~ using~ \jopt{}~ and ~\mean{p}{}}$

\vspace*{0.3in} 
\hrulefill
\vspace*{-0.4in}

\caption{$b$-step $L_p$ approximation of isotonic data, using total monotonicity for $\mathsf{determine\_jvalues}$} 
\label{alg:totallymonotonic}

\end{algorithm}

\begin{algorithm}
\setlength{\Ainindent}{0.1in}
\Ain{0} $\mathsf{procedure~ determine\_jvalues(jvalues,num\_jvalues,c)}$\\

\Ain{0} $\mathsf{num\_jvalues(0)=n}$\\
\Ain{0} $\mathsf{for~ k=1 ~to~ n ~do~ jvalues(0,k)=k}$\\
\Ain{0} $\mathsf{for~ level = 1 ~to~ \lfloor \log_2 (n\!-\!c\!+\!1) \rfloor}$\\
\Ain{1} $\mathsf{   j=jvalues(level\!-\!1,1);~~ next\_j\_index=2}$;~~ k\_index=0\\
\Ain{1} $\mathsf{   i=2^{level}}$\\
\Ain{1} $\mathsf{   while~  next\_j\_index \leq num\_jvalues(level\!-\!1) ~do~}$\\
\Ain{2} $\mathsf{      next\_j = jvalues(level\!-\!1,next\_j\_index)}$\\
\Ain{2} $\mathsf{      if~ (j \geq i) \wedge (\Errj{i,j,c} \leq \Errj{i,next\_j,c}) ~then~}$\\
\Ain{3} $\mathsf{         if~ i+2^{level}>n-c+1 ~then~ \{at~ last~ row,~ eliminate~ next\_j\}}$\\
\Ain{4} $\mathsf{            next\_j\_index = next\_j\_index+1}$\\
\Ain{3} $\mathsf{         else~ \{keep~ this~ j, ~increment~ i,~ j\}}$\\
\Ain{4} $\mathsf{            k\_index=k\_index +1;~~ jvalues(level,k\_index)=j}$\\
\Ain{4} $\mathsf{            j= next\_j;~~ next\_j\_index=next\_j\_index + 1}$\\
\Ain{4} $\mathsf{            i=i+2^{level}}$\\
\Ain{3} $\mathsf{         end~ if}$\\
\Ain{2} $\mathsf{      else~ \{\Errj{i,j,c} > \Errj{i,next\_j,c}, eliminate~ current~ j,~ go~ back~ to~ previous~ i,~ j\}}$\\
\Ain{3} $\mathsf{         if~ i > 2^{level} ~then}$\\
\Ain{4} $\mathsf{             i = i-2^{level};~~ j=jvalues(level,k\_index);~~ k\_index = k\_index-1}$\\
\Ain{3} $\mathsf{         else~ \{at~ first~ row\}}$\\
\Ain{4} $\mathsf{             j=next\_j;~~ next\_j\_index=next\_j\_index+1}$\\
\Ain{3} $\mathsf{         end if}$\\
\Ain{2} $\mathsf{      end~ if}$\\
\Ain{1} $\mathsf{   end~ while}$\\
\Ain{1} $\mathsf{   k\_index = k\_index+1;~~ jvalues(level,k\_index) = j}$\\
\Ain{1} $\mathsf{   num\_jvalues(level)=k\_index}$\\
\Ain{0} $\mathsf{end~ for~ level}$\\
\Ain{0} $\mathsf{end~ determine\_jvalues}$\\

\caption{Reducing the number of relevant $j$ values using SMAWK} \label{alg:SMAWK}

\end{algorithm}

\begin{proposition} \label{prop:SMAWK}
Given $n$ isotonic weighted values \data\ and number of steps $b \leq n$,
Algorithm~\ref{alg:totallymonotonic} finds an optimal $b$-step
$L_2$ approximation in $\Theta(b n)$ time.
\end{proposition}
\textit{Proof:}\,
Since each level halves the number of rows and the number of kept $j$ values is no more than the number of rows, the total number of $j$ values kept over all levels is $O(n)$ and the total time of $\mathsf{determine\_jvalues}$ is $\Theta(n)$.
The time for Algorithm~\ref{alg:totallymonotonic} is linear in the total number of $j$ values considered, so it too is $\Theta(n)$.
$\Box$.

\section{Reduced Isotonic Regression} \label{sec:Reduced}

For arbitrary data, isotonic regressions are somewhat easier to compute than are general
approximations by step functions.
One can use a simple left-right scan where each location is
initially a step and then adjacent steps are merged
whenever they violate the isotonic condition.
This is known as ``pool adjacent violators'', PAV, 
and first appeared in 1955 in Ayer et al.~\cite{Ayeretal1955}.
For $L_2$ it can easily be computed in only $\Theta(n)$ time.

Isotonic regression is a very flexible nonparametric approach to 
many problems. 
However it does have its detractors
due to results with impractically many steps or overfitting. 
Some researchers have instead used approximations with a 
specified number of steps~\cite{HaiminenetalReducedUnimodal,TerziSegmentation}.
To reduce overfitting, Schell and Singh~\cite{SchellSingh} used the approach of
repeatedly merging pairs of adjacent steps whose difference 
had the least statistical significance. 
Haiminen et al.~\cite{HaiminenetalReducedUnimodal} used an approach that
repeatedly combines the adjacent steps that cause a minimum increase in the error.
These greedy (aka myopic) approaches repeatedly make
the choice that seems to be the best at the moment, but
may not produce an optimal reduced isotonic regression.
For example, for all $L_p$, $1 < p \leq \infty$,
given the unweighted values 0, 2, 4, 6, 8, 10, the unique optimal 3-step isotonic regression is 1, 1, 5, 5, 9, 9, 
and the unique optimal 2-step isotonic regression is 2, 2, 2, 8, 8, 8.
Thus the 2-step isotonic regression cannot be obtained by merging steps of the
3-step isotonic regression.

The fastest previous algorithm for optimal $L_2$ reduced isotonic regression is due to Haiminen et al.\ \cite{HaiminenetalReducedUnimodal}, taking $\Theta(n + b m^2)$ time,
where $m$ is the number of pieces in the unconstrained isotonic regression.
As a reminder, we use ``pieces'' to refer to the steps of an unrestricted
isotonic regression and ``steps'' to refer to the steps of a reduced isotonic regression.
Even though often $m \ll n$,  Haiminen et al.\ felt that this may be too slow so they developed the greedy heuristic mentioned above.
Our exact algorithms should be sufficiently fast even for very large problems.

One cannot directly find $b$-step reduced isotonic regression of arbitrary data by using the approaches in Algorithms~\ref{alg:monotonic} and~\ref{alg:totallymonotonic} since it does not have the required monotonic properties.
For example, for unweighted values 7, 8, 0, 6, 9, 10, the optimal $L_2$ 2-step reduced isotonic regression has its first step on the interval \inter{1}{4},
while the optimal first step for the data starting at position 3 is the interval \inter{3}{3}, i.e., 
$4=\jopt{1,2} \not\leq \jopt{3,2}=3$.
Howevever, a critical observation in Haiminen et al.\ \cite{HaiminenetalReducedUnimodal}
is that, given the pieces of an $L_2$ unrestricted isotonic regression, 
the steps of an optimal 
$L_2$ reduced isotonic regression can be formed by merging the pieces.
Each piece becomes a weighted point, where the value of the point is the mean of the piece and the weight of the point is the total weight of the piece.
In the above example, the data would be represented by the 4 weighted points (5,3), (6,1), (9,1), (10,1),
and the first step of a 2-step reduced isotonic regression uses the first two pieces.

Their observation gives a simple algorithm:
find the unrestricted isotonic regression, convert the pieces to weighted points, and then find a $b$-step approximation of these isotonic points.
Haiminen et al.\ used Fisher's algorithm to determine the optimal $b$-step reduced isotonic regression in $\Theta(n+bm^2)$ time,
but Algorithms~\ref{alg:monotonic} and~\ref{alg:totallymonotonic} provide faster solutions.

\begin{theorem}   \label{thm:L2}
Given $n$ weighted values \data\ and number of steps $b$, 
an optimal $L_2$ $b$-step reduced isotonic regression can be found in
$\Theta(n + b m \log m)$ time via Algorithm~\ref{alg:monotonic}, and in $\Theta(n + bm)$ time via
Algorithm~\ref{alg:totallymonotonic},
where $m$ is the number of pieces in the unconstrained $L_2$ isotonic regression.
$\Box$
\end{theorem}

Unfortunately, for $p \neq 2$ the optimal reduced isotonic regression might not be formed from pieces of the unrestricted isotonic regression.
For example, for unweighted values -10, -10, -10, 0, 0, 0, -10, -1, 7, 7, 7, 7, the unique $L_1$ 
unrestricted isotonic regression has pieces 
\inter{1}{3}, \inter{4}{8}, and \inter{9}{12}, with values -10, 0, 7, respectively.
The unique optimal 2-step reduced isotonic regression has 
steps \inter{1}{7} and \inter{8}{12},
with values -10 and 7, which requires cleaving the middle piece.
However, one can determine an approximation by constructing an optimal $b$-step isotonic regression among those restricted to use unbroken pieces of the unrestricted isotonic regression.
By doing so, the problem is now similar to isotonic regression on isotonic data.
An algorithm using this approach to approximate $L_1$ reduced isotonic regression
appears in~\cite{JQReducedIso_IF2012}.
It is more complicated than the $L_2$ case since to determine medians one needs to retain the values in the original pieces, rather than combining them into a single weighted value as can be done for $L_2$.

For $L_\infty$ an optimal $b$-step reduced isotonic regression, and an optimal $b$-step approximation with no isotonic restrictions, can be found in $\Theta(n + \log n \cdot b(1 + \log n/b))$ time~\cite{QLinftyReduced}.
The approaches used there are quite different, unrelated to dynamic programming.

\section{Final Comments}\label{sec:Final}

The thousands of citations to the books by Barlow et al.~\cite{BarlowetalBook} 
and Robertson et al.~\cite{RobertsonWrightDykstra} shows a significant interest in isotonic regression.
Further, this interest is growing as researchers seek to remove parametric assumptions from their modeling.
Similarly, step functions with a constraint on the number of steps arise in a wide range of applications and guises~\cite{FisherHomogeneity,HalimetalFastHistogram,HimbergetalTimeSeries,IoannidisIncreasingHist,JacobetalFuzzySegmentation,MaysterLopezStep,PoosalaetalHistogramSurvey,TerziSegmentation}.
For reduced isotonic regression both aspects are important~\cite{HaiminenetalReducedUnimodal,SalantiUlmReduced,SchellSingh}, using a reduced number of steps to simplify the regression and/or prevent overfitting.

However, researchers used approximations, rather than the optimal answer, due to the slowness of the available algorithms.
The fastest previous algorithm for optimal $L_2$ $b$-step reduced isotonic regression takes $\Theta(n + bm^2)$ time~\cite{HaiminenetalReducedUnimodal}, where $m$ is the number of pieces in the unconstrained isotonic regression.
Algorithm~\ref{alg:monotonic} reduces this to $\Theta(n + b m \log m)$ time,
and the somewhat more complicated Algorithm~\ref{alg:totallymonotonic}
further reduces this to $\Theta(n + b m)$.
Note that the minimal time for optimal $b$-step approximation, with no isotonic restrictions, is a long-standing open question.

Fisher~\cite{FisherHomogeneity} called the $b$-step approximations ``restricted homogenization'', and defined another form of approximation that he
called ``unrestricted homogenization'': given $n$ weighted values \data\ and $b \in \inter{1}{n}$, 
partition the values into $b$ subsets $P_i, i \in \inter{1}{b}$ and assign a value $C_i$
to each $P_i$ so as to minimize
$$
\sum_{i=1}^b \sum_{j \in P_i} w_j |y_j - C_i|^2
$$
among all such partitions.
This is now known as \textit{$k$-means clustering} of 1-dimensional data, for $k = b$.
He noted it could be solved by sorting the values and then finding
the optimal $b$-step approximation, i.e., the optimal $b$-step isotonic
regression of the sorted data.
Thus for 1-dimensional data Algorithm~\ref{alg:monotonic} solves the $k$-means clustering problem in $\Theta(kn \log n)$ time, and
for sorted data Algorithm~\ref{alg:totallymonotonic} reduces this to $\Theta(kn)$.

Finally, an interesting problem is that of selecting the most desirable number of steps.
For reduced isotonic regression,
Schell and Singh~\cite{SchellSingh}, Strobl et al.~\cite{StrobletalCARTreduced} and Haiminen et al.~\cite{HaiminenetalReducedUnimodal} start with an unconstrained isotonic
regression and then repeatedly merge pieces until their criteria are met.
However, Haiminen et al.\ showed that the regression error of their greedy approximation can be nearly twice that of the optimal reduced isotonic regression with the same number of steps.
They believe that 2 is an upper bound on the relative error of their approximation, but that has not been proven, nor have bounds been proven for other approximation schemes.
For $b$-step approximation,
many researchers choose $b$ \textit{a priori} based on considerations
such as storage or access time requirements.
This seems to be especially true in the database community, where $L_2$ $b$-step approximations
are known as ``v-optimal histograms''.

In contrast, the dynamic programming approach generates optimal 
$b$-step reduced isotonic regressions for each value of $b$ as $b$ increases.
One can stop when a criterion is met and always have an optimal result.
However, appropriate stopping criteria for a given application may be somewhat subtle
since they would be applied repeatedly.

\subsubsection*{Acknowledgements}
Research partially supported by NSF grant CDI-1027192 and DOE grant
DE-FC52-08NA28616.
Some of these results were announced in~\cite{JQReducedIso_IF2012}.


\begin{thebibliography}{99}

\bibitem{SMAWK} Aggarwal, A, Klawe, MA, Moran, S, Shor, P and Wilber, R (1987),
 ``Geometric applications of a matrix-searching algorithm'',
 \textit{Algorithmica} 2, pp.~195--208.

\bibitem{Ayeretal1955} Ayer, M, Brunk, HD, Ewing, GM, Reid, WT, and Silverman, E (1955),
  ``An empirical distribution function for sampling with incomplete information'',
  \textit{Annals of Math.\ Stat.}\ 5, pp.~641--647.

\bibitem{BarlowetalBook} Barlow, RE, Bartholomew, DJ, Bremner, JM, and Brunk, HD
  (1972),
  \textit{Statistical Inference Under Order Restrictions: The Theory and Application of Isotonic Regression}, John Wiley.
  
\bibitem{FisherHomogeneity} Fisher, WD (1958),
  ``On grouping for maximum homogeneity'',
  \textit{J.\ Amer.\ Stat.\ Assoc.} 53, pp.~789--798.
   
\bibitem{HaiminenetalReducedUnimodal} 
  Haiminen, N, Gionis, A, and Laasonen, K (2008),
  ``Algorithms for unimodal segmentation with applications to unimodality detection'',
  \textit{Knowl.\ Info.\ Sys.} 14, pp.~39--57.

\bibitem{HalimetalFastHistogram}
  Halim, F, Karras, P, and Yap, RHC (2009),
  ``Fast and effective histogram construction'',
  \textit{Proc.\ Conf.\ Info.\ and Knowl.\ Manag.}, pp.~1167--1176.
  
\bibitem{JQReducedIso_IF2012} 
  Hardwick, J and Stout, QF (2012),
  ``Optimal reduced isotonic reduction'',
  \textit{Proc.\ Interface 2012}, May 2012.
  
\bibitem{HimbergetalTimeSeries} Himberg, J, Korpiaho, K, Mannila, H, Tikanmaki, J
  and Toivonen, H (2001),
  ``Time series segmentation for context recognition in mobile devices'',
  \textit{Int'l.\ Conf.\ Data Mining}, pp.~203--210.
  
\bibitem{IoannidisIncreasingHist}
  Ioannidis, YE (1993),
  ``Universality of serial histograms'',
  \textit{Proc.\ 19th VLDB Conf.}, pp.~256--267.
  
\bibitem{JacobetalFuzzySegmentation}
  Jacob, E, Nair, KNR, and Sasikumar, R (2009),
  ``A fuzzy-driven genetic algorithm for sequence segmentation applied to genomic
  sequences'',
  \textit{Applied Soft Computing} 9, pp.~488--496.
  
\bibitem{MaysterLopezStep}
  Mayster, Y and Lopez, MA (2006),
  ``Approximating a set of points by a step function'',
  \textit{J. Vis.\ Commun.\ Image R.} 17, pp.\ 1178--1189.
  
\bibitem{NiculescuMizilCaruana}
  Niculescu-Mizil, A, and Caruana, R (2005),
  ``Predicting good probabilities with supervised learning'',
  \textit{Proc.\ Int'l.\ Conf.\ Machine Learning} 22, pp.\ 625--632.

\bibitem{PoosalaetalHistogramSurvey}
  Poosala, V, Ioannidis, Y, Haas, P, and Shekita, E (1996),
  ``Improved histograms for selectivity estimation of range predicates'',
  \textit{Proc.\ SIGMOD}, pp.\ 294--305.

\bibitem{RobertsonWrightDykstra} Robertson, T, Wright, FT, and Dykstra, RL (1988),
  \textit{Order Restricted Statistical Inference}, Wiley.
  
\bibitem{SalantiUlmReduced}
  Salanti, G and Ulm, K (2003),
  ``A nonparametric changepoint model for stratifying continuous variables under order
  restrictions and binary outcome'',
  \textit{Stat.\ Methods Med.\ Res.} 12, pp.\ 351--367.

\bibitem{SchellSingh}
  Schell, MJ and Singh, B (1997),
  ``The reduced monotonic regression method'',
  \textit{J.\ Amer.\ Stat.\ Assoc.} 92, pp.\ 128--135.

\bibitem{QLinftyReduced}
  Stout, QF (2014),
 ``An algorithm for $L_\infty$ approximation by a step function'', arXiv 1412.2379

\bibitem{StrobletalCARTreduced}
Strobl, R, Salanti, F, and Ulm, K (2003),
 ``Extension of CART using multiple splits under order restrictions'', Discussion paper, Sonderforschungsbereich 386 der Ludwig-Maximilians-Universitat Munchen, No.\ 364
  
\bibitem{TerziSegmentation} 
  Terzi, E and Tsaparas, P (2006),
  ``Efficient algorithms for sequence segmentation'',
  \textit{Proc.\ 6th SIAM Conf.\ Data Mining}. 

\end{thebibliography}
\end{document}